# The Two-User Deterministic Interference Channel with Rate-Limited Feedback

Alireza Vahid
School of Electrical and Computer Engineering
Cornell University
Email: av292@cornell.edu

A. Salman Avestimehr
School of Electrical and Computer Engineering
Cornell University
Email: avestimehr@ece.cornell.edu

*Abstract*—In this paper we study the effect of *rate-limited* feedback on the sum-rate capacity of the deterministic interference channel. We characterize the sum-rate capacity of this channel in the symmetric case and show that having feedback links can increase the sum-rate capacity by at most the rate of the available feedback. Our proof includes a novel upper-bound on the sum-rate capacity and a set of new achievability strategies.

## I. INTRODUCTION

In many communication scenarios there are feedback links available from the receivers back to the transmitters. It is well-known that feedback does not increase the capacity of discrete-memoryless point-to-point channels [1]. However, feedback can enlarge the capacity region of multi-user networks, even in the two-user memoryless multiple-access channel [2], [3]. Hence, there has been a growing interest in developing feedback strategies and finding upper bounds on the capacity region of networks with feedback, in particular the two-user interference channel (for example see [4]–[8]). Recently in [9], authors have approximated the capacity region of the two-user Gaussian interference channel with feedback and have shown that, quite interestingly, feedback can provide an unbounded gain (as signal-to-noise ratios increase).

Most prior work on interference channel focus on the extreme case in which the feedback link has infinite capacity. However, a more realistic model is one where we have rate-limited feedback from the receivers to the transmitters. Throughout this paper, we will study such a network. As a stepping stone, we focus on the linear deterministic [10] interference channel with rate-limited feedback. The simple linear deterministic model captures the key properties of the Gaussian channel and often provides insights that can lead to approximating the capacity of Gaussian networks [11]–[13].

The main contribution of the paper is the characterization of the sum-rate capacity of the symmetric deterministic interference channel with rate-limited feedback. The sum-rate capacity derived in our work, clearly depicts how we approach the sum-rate capacity of a network with infinte feedback from the sum-rate capacity of a network without any feedback. Moreover, as we will see throughout the paper, there is always a specific value of feedback which will be similar to the case where we have infinite feedback.

Our proof includes a novel upper-bound on the sum-rate capacity and a set of transmission strategies to achieve it. We basically show that the sum-rate capacity cannot be increased by more than the rate of the available feedback. A key property that we have used to prove this result is the linearity of the channel. Hence, an interesting open question for future work is: "Is it in general, *i.e.* without the linearity of the channel, true that the sum-rate capacity of the interference channel can at most be increased by the amount of the available feedback?".

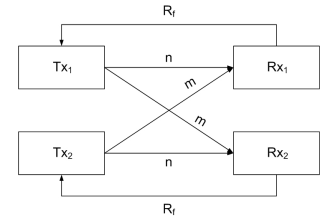

Fig. 1. Two-user interference channel with rate-limited feedback.

The rest of the paper is organized as follows, in Section II we describe our problem formulation and give our main result. Then, in Section III, we propose the capacity achieving schemes. In Section IV, we prove the converse. Section V depicts the sum-rate capacity for the limited feedback case as well as the infinite and the non-feedback cases. Section VI concludes the paper.

## II. PROBLEM SETUP AND THE MAIN RESULT

In this section, we describe the problem formulation and state our main result. We consider the two-user interference network depicted in Figure 1. In this network transmitters 1 and 2, want to communicate with receivers 1 and 2, respectively. We use the deterministic channel model introduced in [10] to model the channels between the tansmitters and the receivers. In this model, there is a non-negative integer associated with each channel which represents its gain. As shown in Figure 1, we consider the symmetric case in which the direct links between the transmitters and the receivers have equal gains (denoted by $n$) and the interfering links have also equal gains (denoted by $m$).

In the deterministic interference channel, we can write the channel input to the transmitter $i$ at time $k$ as $X_i^k = [X_{i1}^k X_{i2}^k \ldots X_{iq}^k]^T$, $i = 1, 2$, such that $X_{i1}^k$ and $X_{iq}^k$ respectively represent the most and the least significant bits of the transmit signal. Also $q$ is the maximum of the channel gains in the network, *i.e.* $q = \max(n, m)$. At each time $k$, the received signal at receiver $i$ ($i = 1, 2$) in our linear model, denoted by $Y_i^k$, is given by

$$Y_1^k = S^{q-n} X_1^k \oplus S^{q-m} X_2^k \\ Y_2^k = S^{q-m} X_1^k \oplus S^{q-n} X_2^k \quad (1)$$

where $S$ is the $q \times q$ shift matrix.

We also assume that there is a noiseless rate-limited, with rate $R_f$, feedback from each receiver to the corresponding transmitter. Here transmitter 1 and transmitter 2 wish to communicate reliably messages $W_1 \in \{1, 2, \ldots, 2^{KR_1}\}$ and $W_2 \in \{1, 2, \ldots, 2^{KR_2}\}$ with receivers 1 and 2, respectively, by $K$ uses of the channels. We say that a rate pair $(R_1, R_2)$ is achievable, if there exists a block encoder at each transmitter (that can causally use the feedback) and a block decoder at each receiver, such that the error probability of decoding the desired message at each receiver goes to zero as the block length $K$ goes to infinity. The maximum sum-rate, *i.e.* $R_1 + R_2$, of all achievable rate pairs is called the sum-rate capacity. The following Theorem is our main result

*Theorem 2.1:* The normalized sum-rate capacity of a deterministic interference channel with rate-limited feedback is given by

$$\frac{C_{\text{sum}}}{n} = \begin{cases} \min(2 - 2\alpha + 2\beta, 2 - \alpha) & \text{for } \alpha \in [0, 0.5] \\ \min(2\alpha + 2\beta, 2 - \alpha) & \text{for } \alpha \in [0.5, \frac{2}{3}] \\ 2 - \alpha & \text{for } \alpha \in [\frac{2}{3}, 1] \\ \alpha & \text{for } \alpha \in [1, 2 + 2\beta] \\ 2 + 2\beta & \text{for } \alpha \in [2 + 2\beta, \infty] \end{cases} \quad (2)$$

where $\beta$ is the normalized feedback rate, i.e $\beta = \frac{R_f}{n}$, and $\alpha$ is the ratio of the gain of the indirect link to that of the direct link, *i.e.* $\alpha = \frac{m}{n}$.

If we compare equation (2) to the sum-rate capacity without feedback [14], [15]

$$\frac{C_{\text{sum}}}{n} = \begin{cases} 2 - 2\alpha & \text{for } \alpha \in [0, 0.5] \\ 2\alpha & \text{for } \alpha \in [0.5, \frac{2}{3}] \\ 2 - \alpha & \text{for } \alpha \in [\frac{2}{3}, 1] \\ \alpha & \text{for } \alpha \in [1, 2] \\ 2 & \text{for } \alpha \in [2, \infty] \end{cases} \quad (3)$$

and the sum-rate capacity with infinite feedback [9]

$$\frac{C_{\text{sum}}}{n} = \begin{cases} 2 - \alpha & \text{for } \alpha \in [0, 1] \\ \alpha & \text{for } \alpha \in [1, \infty] \end{cases}, \quad (4)$$

we note that

- Case 1 ($\alpha \in [0, \frac{1}{2}]$): In this regime the sum-rate capacity is increased by the total amount of feedback rates and saturates at $2 - \alpha$ once the rate of each feedback link is larger than $\frac{m}{2}$.
- Case 2 ($\alpha \in [\frac{1}{2}, \frac{2}{3}]$): In this regime the sum-rate capacity is increased by the total amount of feedback rates and saturates at $2\alpha$ once the rate of each feedback link is larger than $\frac{(2n-3m)}{2}$.
- Case 3 ($\alpha \in [\frac{2}{3}, 2 + 2\beta]$): In this regime feedback does not increase the capacity.

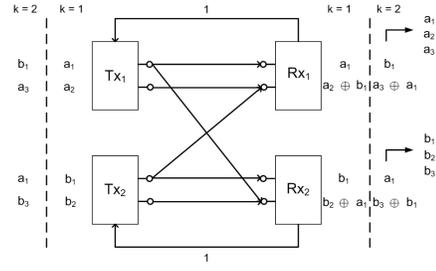

Fig. 2. Scheme to achieve the capacity when $\alpha$ is equal to $\frac{1}{2}$.

- Case 4 ($\alpha \in [2 + 2\beta, \infty]$): In this regime the sum-rate capacity is always increased by the total amount of feedback rates.

## III. TRANSMISSION STRATEGIES

In this section we will provide the transmission strategies to achieve the capacity in Theorem 2.1.

- Case 1 ($\alpha \in [0, \frac{1}{2}]$)

We first start with the example shown in Figure 2. In this example, $n = 2$ and $m = 1$ and $R_f = 1$. We show that each transmitter can send 3 bits to its destination in two time slots and hence achieve a normalized sum-rate of $2 - \alpha = \frac{3}{2}$. In the first time slot, each transmitter transmits two new data bits. Since the channel is symmetric, we will discuss only one of the receivers. Receiver one gets $a_1$ without any trouble. It also receives $a_2 \oplus b_1$. Now through feedback link, receiver one sends $a_2 \oplus b_1$ to transmitter one. In the second time slot each transmitter can decode the MSB from the other transmitter through the feedback and it will transmit it as the MSB in the next time slot, and it will also transmit a new bit on the LSB.

For example, transmitter one transmits $b_1$ as the MSB in second time slot. Now, receiver one has $a_1$, $b_1$, $a_2 \oplus b_1$, and $a_1 \oplus a_3$. Hence, it can decode all $a_1$, $a_2$ and $a_3$. Receiver 2 can also decode $b_1$, $b_2$ and $b_3$. Therefore, we have reached the sum-rate of 6 bits in two time slots, or equally, 3 bits per time slot. Here, we can see that this result is the same as the infinite feedback case. As we will see throughout the paper, we can always find a value of feedback rate that will behave as the infinite feedback case, and as a result having higher feedback capacity than this value cannot help.

This result is what we have proposed in Theorem 2.1. There we mentioned that the capacity when $\alpha$ is less than or equal to $\frac{1}{2}$ is $\min(2 - 2\alpha + 2\beta, 2 - \alpha)$. In the above example, $\alpha$ is equal to $\frac{1}{2}$ and $\beta$ is also equal to $\frac{1}{2}$. Therefore, from Theorem 2.1, we have the capacity $2 \times (2 - \alpha) = 3$ for this example.

Now that we have seen the strategy through a simple example, we will give a scheme for the general case where $\alpha \leq \frac{1}{2}$.

As shown in Figure 3, in the first time slot, *i.e.* $k = 1$, each transmitter $i$ sends $n - m$ bits, such that it does not cause interference at the receivers. We also send $R_f$ new bits using the top signal levels, *i.e.* $\Delta_{i1}$ where the first index indicates the transmitter and the second index indicates the time slot. These bits will interfere at the receivers. Since the number of interfered levels are $R_f$, receivers will send them back to the transmitters, and each transmitter can remove its own bits and decode the interfering signals. In the second time slot, we repeat the same scheme. However, we also send the $R_f$ interfering bits which has been recovered from the interfered signal, *i.e.* $\Delta_{11}$ and $\Delta_{21}$. With this scheme, each receiver will use these bits to remove the interference in the first time slot. Moreover, the interference that these bits create is not important, because the receivers already know them and can cancel them out. The same scheme can be done for the coming time slots. Through this transmission strategy we have achieved the rate $n - m + R_f$ at each transmitter per time slot, which is the same as what we have proposed in Theorem 2.1.

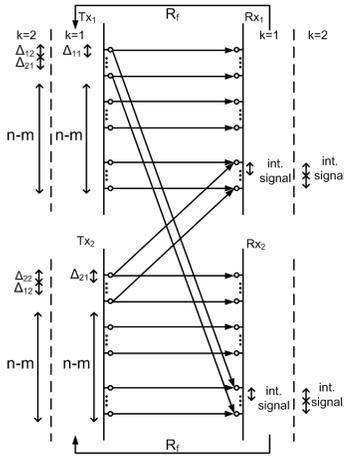

Fig. 3. Transmission strategy for Case 1 ($\alpha \leq \frac{1}{2}$).

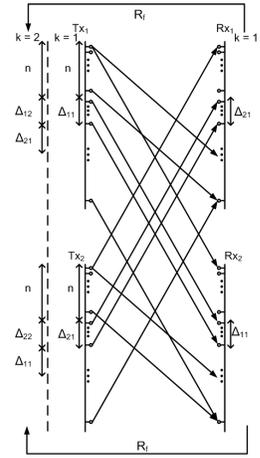

Fig. 5. Transmission strategy for Case 4 ($\alpha \geq 2 + 2\beta$).

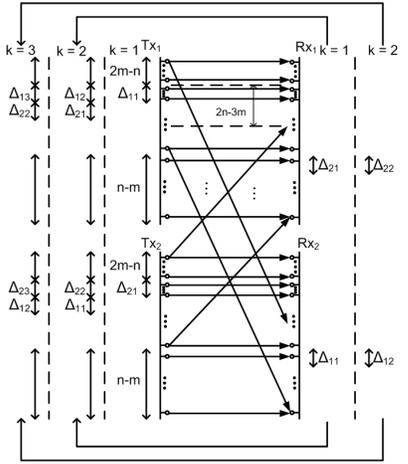

Fig. 4. Transmission strategy for Case 2 ($\alpha \in [\frac{1}{2}, \frac{2}{3}]$).

• Case 2 ($\alpha \in [\frac{1}{2}, \frac{2}{3}]$)

In this case, each transmitter $i$ transmits $2m-n$ bits starting from the MSB and $n-m$ bits on less significant bits, in the first time slot, as shown in Figure 4. Each transmitter also transmits $R_f$ bits as depicted, *i.e.* $\Delta_{11}$ and $\Delta_{21}$. Consider receiver 1, it receives $m$ bits from transmitter 1 and $R_f$ bits from transmitter 2. Now, it will send the $R_f$ bits from transmitter 2 to transmitter 1 through the feedback link. In the second time slot, we use the same strategy as in the first time slot and we also need to send $R_f$ bits received by transmitter 1 through feedback link, *i.e.* $\Delta_{21}$. To avoid interference at the receiver, we should have $R_f$ less than or equal to $\frac{(2n-3m)}{2}$. However, looking at the capacity of the infinite feedback case in [9], we can easily conclude that this is exactly the amount of feedback which will result in a behavior similar to the infinite feedback case. We do exactly the same scheme for the coming time slots. With this scheme we have achieved the normalized sum-rate $2\alpha + 2\beta$.

• Case 3 ($\alpha \in [\frac{2}{3}, 2]$)

As we noted just after Theorem 2.1, for $\alpha$ between $\frac{2}{3}$ and 2, the capacity cannot be increased with feedback, therefore we can simply use the scheme proposed in [12] for this interval.

• Case 4 ($\alpha$ greater than 2)

First assume that $\alpha$ is greater than $2 + 2\beta$, we use the following method: in the first time slot, each transmitter sends $n + R_f$ new data bits. Receiver 1, receives $n$ bits from transmitter 1 and $n + R_f$ bits from transmitter two. Receiver 1 uses the feedback link to send the $R_f$ bits from transmitter 2, *i.e.* the bits that receiver 2 does not have access to in the first time slot ($\Delta_{21}$), to transmitter 1. In the second time slot, tramitter 1 repeats this scheme and also sends the $R_f$ bits that it has received through feedback. We use exacatly the same strategy for transmitter and receiver 2. In the following time slots, we repeat the same scheme as in the second time slot. The method is depicted in Figure 5. At the receivers, we have only depicted the bits that will be sent back to the transmitters through feedback. With this scheme we achieve the sum-rate $2 + 2\beta$, which is the same sum-rate suggested in Theorem 2.1. Now with the scheme just described, it can be easily seen that when $\alpha$ is smaller than $2 + 2\beta$ and grater than 2, the sum-rate capacity is equal to the case where we have infinite feedback.

## IV. CONVERSE

• Case 1 ($\alpha \leq \frac{1}{2}$)

In order to prove the converse in this case, we first prove the following lemma, which upper bounds $H(P_1^K, V_2^K | V_1^K, Q_1^K)$

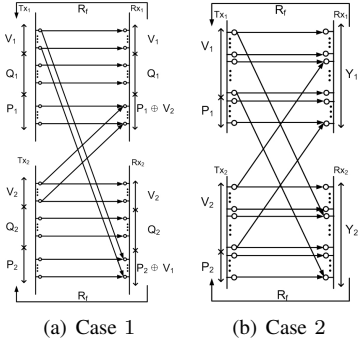

(a) Case 1  (b) Case 2

Fig. 6.

and $H(P_2^K, V_1^K | V_2^K, Q_2^K)$. Here, $V_1^K$ and $V_2^K$ are received by both receivers. $P_1^K$ and $P_2^K$ will be interfered with $V_2^K$ and $V_1^K$ at the receivers, respectively. $Q_1^K$ and $Q_2^K$ are received by only one receiver without any interference, see Figure 6(a).

*Lemma 4.1:* For the linear network described above we have

$$H(P_1^K, V_2^K | V_1^K, Q_1^K) \leq \alpha n K + 1 + K p_e^K$$
$$H(P_2^K, V_1^K | V_2^K, Q_2^K) \leq \alpha n K + 1 + K p_e^K \quad (5)$$

where $p_e^K$ is the decoding error probability.

*Proof:* Given $Y_i^K$, receiver $i$ should be able to decode $X_i^K$ (with high probability as $K \to \infty$), $i = 1, 2$. Therefore, by using the Fano's inequality we have

$$H(P_1^K, V_2^K | Y_1^K) = H(P_1^K | Y_1^K) + H(V_2^K | Y_1^K, P_1^K)$$
$$= H(P_1^K | Y_1^K) \overset{(Fano)}{\leq} 1 + K p_e^K$$
$$H(P_2^K, V_1^K | Y_2^K) = H(P_2^K | Y_2^K) + H(V_1^K | Y_2^K, P_2^K) \quad (6)$$
$$= H(P_1^K | Y_1^K) \overset{(Fano)}{\leq} 1 + K p_e^K$$

Note that $H(V_2^K | Y_1^K, P_1^K) = H(V_1^K | Y_2^K, P_2^K) = 0$, due to the linearity of the channel. Next, we can write

$$H(P_1^K, V_2^K | V_1^K, Q_1^K, P_1^K \oplus V_2^K) \leq 1 + K p_e^K$$
$$H(P_2^K, V_1^K | V_2^K, Q_2^K, P_2^K \oplus V_1^K) \leq 1 + K p_e^K \quad (7)$$

since the terms $P_1^K \oplus V_2^K$ and $P_2^K \oplus V_1^K$ have at most $\alpha n K$ bits each, therefore we have

$$H(P_1^K, V_2^K | V_1^K, Q_1^K) \leq \alpha n K + 1 + K p_e^K$$
$$H(P_2^K, V_1^K | V_2^K, Q_2^K) \leq \alpha n K + 1 + K p_e^K \quad (8)$$

∎

We will also need the next lemma in our proof.

*Lemma 4.2:* $I(X_1^K; X_2^K) \leq 2K R_f$

*Proof:*
We can write

$$I(X_1^K; X_2^K) \leq I(W_1, F_1^K; W_2, F_2^K)$$
$$= I(W_1; W_2) + I(W_1; F_2^K | W_2)$$
$$+ I(F_1^K; W_2 | W_1) + I(F_1^K; F_2^K | W_1, W_2) \quad (9)$$
$$\leq 2K R_f$$

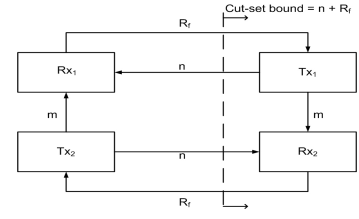

Fig. 7. Cut-set bound when $\alpha \geq 2$.

We know that $W_1$ is independent of $W_2$, therefore: $I(W_1; W_2) = 0$. Moreover given $W_1$ and $W_2$, we can construct everything, as a result $I(F_1^K; F_2^K | W_1, W_2) = 0$. Now since $F_1^K$ and $F_2^K$ are at most $K R_f$ bits each, the proof is complete. ∎

Before proceeding to the proof, we have to note the following inequality which will be used later

$$H(W_1, W_2)$$
$$= I(W_1, W_2; Y_1^K, Y_2^K) + H(W_1, W_2 | Y_1^K, Y_2^K)$$
$$\overset{(a)}{\leq} I(X_1^K, X_2^K; Y_1^K, Y_2^K) + H(W_1, W_2 | Y_1^K, Y_2^K) \quad (10)$$
$$\overset{(b)}{\leq} H(X_1^K, X_2^K) + 2 + 2K p_e^K$$

where (a) follows from data processing and (b) from Fano. We are now ready to prove the converse in this case

$$K(R_1 + R_2) = H(W_1, W_2)$$
$$\leq H(X_2^K) + H(X_1^K | X_2^K) + 2 + 2K p_e^K$$
$$= H(V_2^K) + H(Q_2^K | V_2^K) + H(P_2^K | Q_2^K, V_2^K)$$
$$+ H(V_1^K | V_2^K, Q_2^K, P_2^K) + H(Q_1^K | V_1^K, V_2^K, Q_2^K, P_2^K)$$
$$+ H(P_1^K | V_1^K, V_2^K, Q_1^K, Q_2^K, P_2^K) + 2 + 2K p_e^K$$
$$\leq H(V_2^K | V_1^K, Q_1^K) + H(P_1^K | V_1^K, V_2^K, Q_1^K)$$
$$+ H(P_2^K | V_2^K, Q_2^K) + H(V_1^K | V_2^K, P_2^K, Q_2^K)$$
$$+ H(Q_2^K) + H(Q_1^K) + I(V_2^K; V_1^K, Q_1^K) + 2 + 2K p_e^K$$
$$= H(P_1^K, V_2^K | V_1^K, Q_1^K) + H(P_2^K, V_1^K | V_2^K, Q_2^K)$$
$$+ I(V_2^K; V_1^K, Q_1^K) + H(Q_2^K) + H(Q_1^K) + 2 + 2K p_e^K$$
$$\leq 2Kn - 2\alpha Kn + 2K R_f + 4 + 4K p_e^K$$
(11)

where the last step is true, since

1) By Lemma 4.1, $H(P_1^K, V_2^K | V_1^K, Q_1^K) + H(P_2^K, V_1^K | V_2^K, Q_2^K) \leq 2\alpha n K + 2 + 2K p_e^K$.
2) By Lemma 4.2, $I(V_2^K; V_1^K, Q_1^K) \leq I(X_1^K; X_2^K) \leq 2K R_f$.
3) $H(Q_1^K) + H(Q_2^K) \leq 2Kn - 4\alpha Kn$, as each is at most $(1 - 2\alpha)Kn$ bits.

The converse can be obtained by dividing both sides by $K$ and letting $K \to \infty$.

• Case 2 ($\alpha \in [0.5, 2/3]$)

This case is depicted in Figure 6(b). Similar to the previous case, we can write

$$\begin{aligned}
K(R_1 + R_2) &= H(W_1, W_2) \\
&\leq H(X_2^K) + H(X_1^K|X_2^K) + 2 + 2Kp_e^K \\
&= H(V_2^K) + H(P_2^k|V_2^K) + H(V_1^K|V_2^K, P_2^K) \\
&\quad + H(P_1^K|V_1^K, V_2^K, P_2^K) + 2 + 2Kp_e^K \\
&\leq H(V_2^K|V_1^K) + H(P_1^K|V_1^K, V_2^K) \\
&\quad + H(P_2^K|V_2^K) + H(V_1^K|V_2^K, P_2^K) \\
&\quad + I(V_2^K; V_1^K) + 2 + 2Kp_e^K \\
&= H(P_1^K, V_2^K|V_1^K) + H(P_2^K, V_1^K|V_2^K) \\
&\quad + I(V_2^K; V_1^K) + 2 + 2Kp_e^K \\
&\leq 2\alpha Kn + 2KR_f + 4 + 4Kp_e^K
\end{aligned} \quad (12)$$

where the last step is true, since

1) Similar to Lemma 4.1, we can show that
$$\begin{aligned}
H(P_1^K, V_2^K|V_1^K) &\leq \alpha n K + 1 + Kp_e^K \\
H(P_2^K, V_1^K|V_2^K) &\leq \alpha n K + 1 + Kp_e^K
\end{aligned} \quad (13)$$

2) By Lemma 4.2, $I(V_2^K; V_1^K) \leq I(X_1^K; X_2^K) \leq 2KR_f$.

The converse can be obtained by dividing both sides by $K$ and letting $K \to \infty$.

- Case 3 ($\alpha \in [\frac{2}{3}, 2]$)

For this interval feedback does not increase the capacity and as a result we can use the proof in [9] or [12].

- Case 4 ($\alpha \geq 2$)

In this case we prove the converse by using a cut-set bound, similar to the one used in [8]. The linear network is shown in Figure 7, consider the cut (dashed line) which separates each transmitter from its receiver (note that we have interchanged the place of one of the transmitter-receiver pairs in this figure). Since the sum of the capacities of the links going from each side of the cut to the other side is equal to $n + R_f$, the sum-rate capacity is upper-bounded by $2n + 2R_f$. As a result the normalized sum-rate is upper bounded by $2 + 2\beta$. This completes the proof of Theorem 2.1.

## V. PLOTS

In Figure 8, we have plotted the sum-rate capacity of the deterministic interference channel for $R_f = 0.125 \times n$, as well as the extreme cases, i.e. infinite feedback and no feedback cases. As $R_f$ increases the curve approaches the infinite feedback case. Another case that can be considered, is where we have variable feedback. However, we just considered the case where $R_f$ is constant. We should pay attention that we are focusing on the sum-rate and therefore our curve starts at 2 for $\alpha = 0$.

## VI. CONCLUSION

We studied the deterministic interference channel with rate-limited feedback. We fully characterized the sum-rate capacity of the symmetric deterministic interference channel with rate-limited feedback. We developed a new outer bound which shows that the sum-rate capacity cannot be increased by more than the total rate of available feedback. Our proof of the converse relies on the linearity of the channel. An interesting open problem is to understand whether the same is true in general (*i.e.* when the channels are not necessarily linear). The next step is to generalize our converse and achievability strategies to the Gaussian networks.

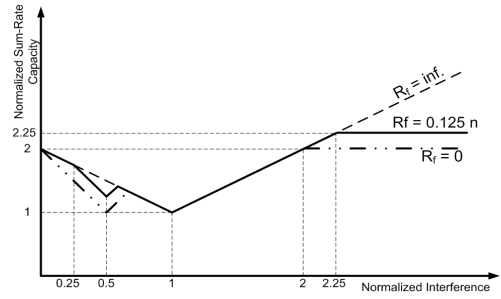

Fig. 8. Proposed capacity as well as infinite feedback and non feedback scenarios.

## VII. ACKNOWLEDGEMENT

This work is supported in part by NSF CAREER award 0953117.

## REFERENCES

[1] C. E. Shannon, "The Zero Error Capacity of a Noisy Channel," *IRE Transactions on Information Theory*, Sept. 1956.
[2] N. T. Gaarder and J. K. Wolf, "The capacity region of a multipleaccess discrete memoryless channenl can increase with feedback," *IEEE Transactions on Information Theory*, Jan. 1975.
[3] L. H. Ozarow, "The capacity of the white Gaussian multiple access channel with feedback," *IEEE Transactions on Information Theory*, July 1984.
[4] G. Kramer, "Feedback Strategies for White Gaussian Interference Networks," *IEEE Transactions on Information Theory*, June 2002.
[5] G. Kramer, "Correction to Feedback Strategies for White Gaussian Interference Networks, and a Capacity Theorem for Gaussian Interference Channels with Feedback," *IEEE Transactions on Information Theory*, June 2004.
[6] M. Gastpar and G. Kramer, "On Noisy Feedback for Interference Channels," *In Proceedings Asilomar Conference on Signals, Systems and Computers*, Oct. 2006.
[7] J. Jiang, Y. Xin, and H. K. Garg, "Discrete Memoryless Interference Channels with Feedback," *CISS 41st Annual Conference*, Mar. 2007.
[8] A. Sahai, V. Aggarwal, M. Yuksel, and A. Sabharwal, "On Channel Output Feedback in Deterministic Interference Channels," *In Proceedings IEEE ITW*, Oct. 2009.
[9] C. Suh, and D. Tse, "Symmetric feedback capacity of the Gaussian interference channel to within one bit," *in proceeding IEEE International Symposium on Information Theory (ISIT)*, 2009.
[10] S. Avestimehr, S. Diggavi, and D. Tse, "A deterministic approach to wireless relay networks," *Proceedings of Allerton Conference on Communication, Control, and Computing*, Sept. 2007.
[11] S. Avestimehr, S. Diggavi, and D. Tse, "Wireless Network Information Flow: A Detreministic Approach," *submitted to IEEE Transactions on Information Theory*, 2009. Available online at http://arxiv.org/abs/0906.5394.
[12] G. Bresler, and David Tse, "The Two-User Gaussian Interference Channel: A Deterministic View," *European Transactions in Telecommunications*, June 2008.
[13] S. Avestimehr, A. Sezgin, and D. Tse, "Capacity of the Two Way Relay Channel within a Constant Gap," *European Transactions on Telecommunications*, to appear.
[14] A. El Gamal, M. Costa, "The capacity region of a class of deterministic interference channels", *IEEE Transactions on Information Theory*, Mar. 1982.
[15] R. Etkin, D. Tse, and H. Wang, "Gaussian interference channel capacity to within one bit," *IEEE Transactions on Information Theory*, Dec. 2008.